\documentclass{aa}

\usepackage{graphicx}
\usepackage{txfonts}
\begin{document}

\title{Photometric and spectroscopic evidence for a dense ring system around Centaur Chariklo\thanks{Partially based on observations collected at the European Organisation for Astronomical Research in the Southern Hemisphere, Chile. DDT 291.C-5035(A). Based on observations carried out at the Complejo Astron\'omico El Leoncito, which is operated under agreement between the Consejo Nacional de Investigaciones Cient\'{i}ficas y T\'ecnicas de la Rep\'ublica Argentina and the National Universities of La Plata, C\'ordoba, and San Juan.}}


   \author{R. Duffard  \inst{1}
    \and
    N. Pinilla-Alonso \inst{2}
          \and
          J.L. Ortiz\inst{1}
    \and
    A. Alvarez-Candal \inst{3}
    \and
    B. Sicardy \inst{4}
    \and
    P. Santos-Sanz \inst{1}
    \and
    N. Morales \inst{1}
    \and
    C. Colazo \inst{5}
    \and
    E. Fern\'andez-Valenzuela \inst{1}
    \and
    F. Braga-Ribas \inst{3}
          }

   \institute{Instituto de Astrofisica de Andalucia - CSIC. Glorieta de la Astronom\'{i}a s/n. Granada. 18008. Spain\\
              \email{duffard@iaa.es}
         \and
          Department of Earth and Planetary Sciences, University of Tennessee, Knoxville, TN, 37996-1410, USA
    \and
    Observatorio Nacional de Rio de Janeiro, Rio de Janeiro, Brazil.
    \and
    LESIA-Observatoire de Paris, CNRS, UPMC Univ. Paris 6, Univ. Paris-Diderot, 5 place J. Janssen, 92195 Meudon Cedex,
France.
    \and
    Observatorio Astron\'omico, Universidad Nacional de C\'ordoba, Laprida 854, C\'ordoba 5000, Argentina.
             }

   \date{  -----}


  \abstract
{A stellar occultation observed on 3rd June 2013 revealed the presence of two dense and narrow rings separated by a small gap around the Centaur object (10199) Chariklo. The composition of these rings is not known. We suspect that water ice is present in the rings, as is the case for Saturn and other rings around the giant planets.}
{In this work, we aim to determine if the variability in the absolute magnitude of Chariklo and the temporal variation of the spectral ice feature, even when it disappeared in 2007, can be explained by an icy ring system whose aspect angle changes with time.}
{We explained the variations on the absolute magnitude of Chariklo and its ring by modeling the light reflected by a system as the one described above.  Using X-Shooter at VLT, we obtained a new reflectance spectra.  We compared this new set of data with the ones available in the literature. We showed how the water ice feature is visible in 2013 in accordance with the ring configuration, which had an opening angle of nearly 34$^o$ in 2013. Finally, we also used models of light scattering to fit the visible and near-infrared spectra that shows different characteristics to obtain information on the composition of Chariklo and its rings.}
{We showed that absolute photometry of Chariklo from the literature and new photometric data that we obtained in 2013 can be explained by a ring of particles whose opening angle changes as a function of time. We used the two possible pole solutions for the ring system and found that only one of them, $\alpha$=151.30$\pm0.5$, $\delta=41.48\pm0.2$ $^o$ ($\lambda=137.9\pm0.5$, $\beta=27.7\pm0.2$ $^o$), provides the right variation of the aspect angle with time to explain the photometry, whereas the other possible pole solution fails to explain the photometry. From spectral modeling, we derived the composition of the Chariklo surface and that of the rings using the result on the pole solution.  Chariklo surface is composed with about 60\% of amorphous carbon, 30\% of silicates and 10\% of organics; no water ice was found on the surface.  The ring, on the other hand, contains 20\% of water ice,  40-70\% of silicates, and 10-30\% of tholins and small quantities of amorphous carbon. }
{}

   \maketitle
%

\section{Introduction}

Centaurs are thought to be objects that originated in the trans-Neptunian
region and that are currently in a transition phase with unstable orbits
lying between Jupiter and Neptune and dynamical life-times on the order of
10 Myears \citep{Horner2004}. They can become short period comets, and several Centaurs have already shown cometary-like activity. The first
object of this class with cometary activity ever discovered was (2060)
Chiron \citep{Tholen1988}. Some time after its discovery it experienced a brightness outburst, which
developed a coma, and showed typical cometary behavior. Another relevant
object among the Centaurs is (10199) Chariklo, which appears to be the
largest object in this population with a diameter of 248$\pm$18 km
\citep{Fornasier2013, Duffard2014}, in contrast to Chiron, which has never shown
any traces of cometary activity. Vast photometric monitoring along a large
time span has been made since its discovery in 1997 and tries to identify
possible outbursts, but no such potential outbursts have been identified
thus far (e.g. \citealt{Belskaya2010}).

Recently, the observation of a multi-chord stellar occultation reported in \cite{Braga2014} revealed the presence of two dense rings
around Chariklo with widths of about 7 and 4km, optical depths of 0.4 and
0.06, and orbital radii of 391 and 405km, respectively.

In the compilation of the photometry by \cite{Belskaya2010}, the
absolute magnitude of Chariklo increased with time (Chariklo's brightness
decreased) from 6.8 in 1998 to 7.4 in 2009, and this has been interpreted by
some authors as a possible hint for a decreasing comet-like activity in this
body \citep{Guilbert2009b, Belskaya2010}, which seemed reasonable because
several other centaurs are known to have cometary-like activity
\citep{Jewitt2009}, and not only Chiron. Nevertheless,  \cite{Guilbert2011}  showed with independent arguments that cometary activity for Chariklo is currently not possible, unless an additional energy source is provided to the object (through an impact for example).

On the other hand, several independent spectroscopic studies of Chariklo
reported the detection of water ice absorption bands located at 1.5 and 2.0
$\mu$m (e.g. \citealt{BrownRobert1998, BrownMichael1998, Dotto2003}). Later in time, observations
with higher S/N ratio show an albedo that has not signs of the presence of water ice
\citep{Guilbert2009a, Guilbert2009b}. In this work, the authors tried to explain the observed spectral variations in terms of surface heterogeneity, which was a plausible explanation at the moment.

However, in light of the recent discovery by \cite{Braga2014}, there is another option to explain the photometry: 
Chariklo's dimming and the overall photometric behavior of Chariklo is related to the change in the aspect angle of a bright and dense ring system around Chariklo.

A dense ring system is also a natural source that could explain the changes in the depth of the water
ice feature observed in Chariklo's spectrum, as in Saturn's ring \citep{Hedman2013}. 
Moreover, there is one spectrum that shows no indication of water ice, which suggests
that all the water ice detected on Chariklo would not be on Chariklo's
surface but on its rings. One of the ring pole positions derived by
\cite{Braga2014} implies that the rings were seen edge-on in 2007-2008 when no water ice spectral features were observed in Chariklo
\citep{Guilbert2011}. 

Our interpretation also predicts that water ice features should already be
detectable in 2013. To confirm that, we present spectroscopic
observations taken with the 8m Very Large Telescope (VLT) using X-Shooter here, that clearly show the
reappearance of the water ice spectral features in the NIR. 

In summary, \cite{Braga2014} show the detection of the ring system through stellar occultation, and we present evidence here for the ring system
based on photometry and spectroscopy. We also derive additional constraints
to the rings properties that are not directly obtained from the occultation.
In section 2, the new observational data are presented, and the analysis of the data is presented in section 3. Finally, the discussion is shown
in section 4 and conclusions in section 5.


   \begin{figure*}
   \centering
    \includegraphics[width=18cm]{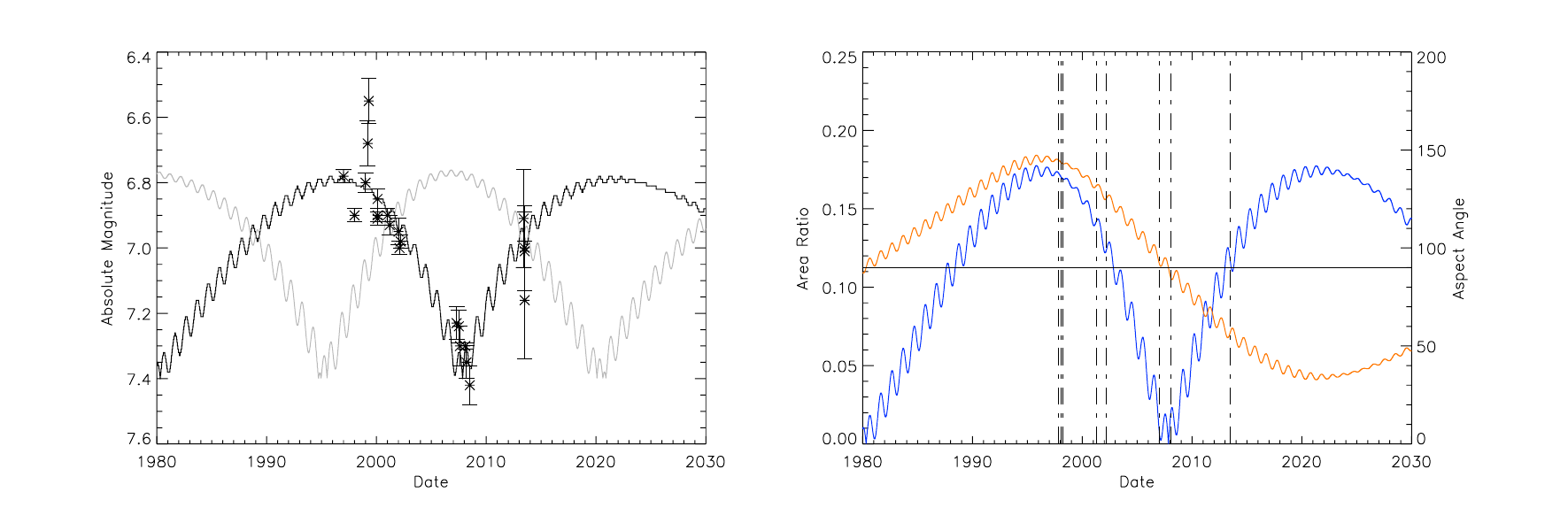}
\caption{Left panel: Variation of the H$_{V}$ during time. Data taken
from \cite{Belskaya2010} and our own observations. The solid
curve represents the model described by Eq. (1). It is fitted to the data, assuming the ring pole position
mentioned in the text ($\lambda=138$, $\beta=28$). We assume that Chariklo's rotation axis is aligned with ring pole. The grey line is the model assuming the other, wrong, pole position.
The right panel shows the variation of the aspect angle with time (orange line, scale at right). As can be seen, the
aspect angle is 90$^o$ in the period 2007-2008. The variation of the area ratio (blue line, scale at left) is also presented in this panel. Vertical lines are the dates when reflectance spectra were taken. The area of Chariklo was considered to be the one of an ellipsoid with axis a=122km, b=122km, and c=117km. The area of the rings were calculated with the parameters taken from \cite{Braga2014}.   }
              \label{Fig1}
    \end{figure*}

\section{New observational data}

With the intention of obtaining new values for the absolute magnitude of
Chariklo, photometric observations were made with different telescopes in 2013. As was mentioned, a reflectance spectra in the vis+nir
range was also obtained with the X-Shoother at VLT.

\subsection{Imaging}

Observations with the Cerro Burek 0.45m Astrograph for the Southern Hemisphere (ASH) telescope in Argentina, the San
Pedro de Atacama 0.4m ASH2 telescope in Chile, and the 1.54m Bosque Alegre
telescope in Argentina were made in different dates through May and June
2013. The exact observing dates at each telescope and other relevant information are summarized in table 1.
The observations consisted of CCD images taken with different cameras. Average seeing was 2 arcsec at ASH, 2.5 arcsec at
ASH2, and 2.1 arcsec at Bosque Alegre. The images were acquired with no
filter to maximize the signal-to-noise ratio. Synthetic aperture
photometry was obtained using aperture diameters that are twice as large as the seeing. Because the sensitivity of the CCD cameras peaked in the R
band, absolute photometry was made using the R magnitudes of UCAC2 reference
stars of similar color to that of Chariklo. This yields magnitudes with an
uncertainty of around 0.15 mag. We used large sets of images to
average out possible rotational variability of Chariklo.  The R magnitudes obtained at each telescope were corrected for
geocentric ($\Delta$) and heliocentric (r) distance by applying
$-5 log(r \Delta)$, where r and $\Delta$ are in astronomical units. The R
magnitudes were translated into V magnitudes by using the known V-R color of
Chariklo (0.48$\pm$0.05), and phase angle corrections were also applied with a 0.06 mag/degree slope parameter
\citep{Belskaya2010}. The absolute magnitude obtained in June 2013 are
7.01$\pm$0.12 mag,  7.00$\pm$0.13 mag, and 7.16$\pm$0.18 mag for ASH2, ASH,
and Bosque Alegre, respectively. For May 2013 observations at ASH gives
H$_V$=6.91$\pm$0.15 mag. These new results are also shown in figure 1 with
the compilation of the literature value. The explanation on how the model was obtained is given below
in Section 3.2.

\subsection{Chariklo reflectance spectra}

We obtained a reflectance spectrum of Chariklo ranging from 0.5 to 2.3 $\mu$m using the
X-Shooter\footnote{\tt http://www.eso.org/sci/facilities/paranal/instruments/xshooter/}
spectrograph located in the Cassegrain focus of unit 2 at the VLT. The X-Shooter is an
{\it echelle} spectrograph that can simultaneously record all the spectral range
by means of two dichroics that split the incoming beam from the telescope and sends it
to three different arms: UVB ($\approx0.3-0.5~\mu$m), VIS ($\approx0.5-1.0~\mu$m),
and NIR ($\approx 1-2.4~\mu$m).

Chariklo was observed on the night of August 3rd, 2013. We used the SLIT mode. The slits chosen were 1.0,
0.9 and 0.9 arcsecs for the UVB, VIS, and NIR arms respectively, which yields a resolving power of about 5000 per arm. To eliminate the sky contribution from the NIR frames we nodded the telescope.
Unfortunately, X-Shooter does not yet allow nodding only in the NIR arm. We
also observed a star to be used as both a telluric and solar analog
(HD144585). All data were reduced using the X-Shooter pipeline by following
the procedure described in \cite{Alvarez2011}, which includes flat-fielding,
wavelength calibration, and merging of different orders. The data were
wavelength and spatially calibrated by creating a two-dimensional wave map, which is necessary because of the curvature of the Echelle orders. We then extracted the one-dimensional spectra using IRAF, which proved better than using the
one-dimensional spectra provided by the pipeline, and we divided those of
Chariklo by the corresponding star. Finally, a median filter was applied to
remove remaining bad pixels. In figure 2, we show the spectrum of Chariklo
from 0.5 up to 2.3 $\mu$m, as obtained in August 2013 with X-Shooter.

  \begin{figure}
   \centering
   \includegraphics[width=8cm]{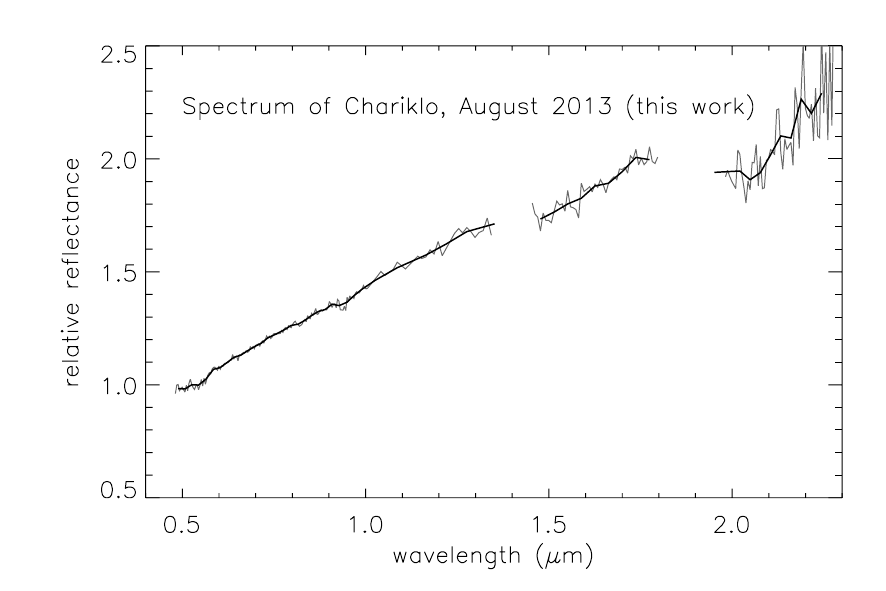}
      \caption{ Reflectance spectra of Chariklo obtained in August 2013 using the X-Shooter at VLT.
              }
        \label{Fig2}
   \end{figure}

\begin{table}
\caption{Observational conditions for the images for obtaining the reported absolute magnitudes and for the new Chariklo spectra obtained with X-Shooter at VLT.}             
\label{table:1}      
\centering                          
\begin{tabular}{c c c c}        
\hline
\multicolumn{4}{c}{Photometry}\\
\hline
Object          &  Telescope      &  exptime (s)   &  Filter            \\
\hline
Chariklo       &  ASH               &  300       &  Luminance    \\
Chariklo       &  ASH2             &  300       & Luminance      \\
Chariklo       &  Bosque Alegre & 300    &  Clear                \\
\hline
\hline
\multicolumn{4}{c}{Near-infrared spectra}\\
\hline
Object   & exptime (s)  &airmass        &    arm     \\
\hline
Chariklo      & $4\times530$&1.035-1.054 & UVB  \\
              & $4\times560$    &            & VIS  \\
              & $4\times600$    &            & NIR  \\
\hline
HD144585 & $2\times0.8$ &1.022-1.023    & All   \\
\hline                                   
\end{tabular}
\end{table}

\section{Analysis of the data}

\subsection {Analysis of the spectrum}

The spectrum of Chariklo taken in 2013 shows a red slope in the visible of
9.375\%/1000$\AA$ from 0.55 to 0.75 $\mu$m. This slope is typically related with the
presence of complex organics on the surface of the body, although it has also
been related in the outer solar system to the presence of amorphous
silicates such as the reddish slope of trojans asteroids \citep{Emery2011}. It also shows two clear absorption bands centered at 1.5 and 2 $\mu$m
typical of water ice. To study the variation on the reflectance from
Chariklo's surface, we compared the parameters of this spectrum (spectral slope in the visible and 2.0 $\mu$m band depth) with others in the literature (see table 2). However, this comparison has to be taken with care. Nowadays,
observational techniques of Centaurs and trans-Neptunian objects are pretty
standard, and solar analogues are widely used to remove the signature of the
Sun in the reflected light from the body. However, this was not always done
in the 90s, when \cite{BrownRobert1998} used the spectrum of a C-type asteroid and
\cite{BrownMichael1998} used the black body function of an A2-type star divided by
that of the sun as seen in the library we are using. Both methods are acceptable but they can affect the slope
of the continuum.

For our study, visible slopes for all the spectra (the new spectrum presented here and those from the literature) were computed using a linear fit to the data between 0.55 and 0.75 $\mu$m. After normalizing all of them at 0.55 $\mu$m.
To count for the error, several fits were randomly run by removing points from
the input data; the slope was chosen as the average value, while the standard
deviation was chosen as the error.

Band depths (\%) were computed (also for all the existing spectra) by defining a linear continuum between 1.75
and 2.2 $\mu$m and then by measuring the reflectance at 2.0 $\mu$m relative to
the one at 1.75 $\mu$m. The determination of the band depth has the problem
that the band is near a large telluric absorption, which leaves considerable
residuals after correction. This might have some incidence on the measured
values.

\subsection{Analysis of the photometric data}

To make an interpretation of the photometry compiled by \cite{Belskaya2010}
and our new photometry during 2013, we built a model of the reflected light
by the main body and a model of the reflected light by the rings. For the main body, we obtained the flux by
multiplying the projected area of an ellipsoid by the geometric albedo of
Chariklo.

The total flux density $F_{tot}$ coming from both Chariklo and its rings is then given by

\begin{equation}
$$ \frac{ F_{tot}}{F_{Sun}} = A_p  p_v f(\alpha) + p^{R1}_v 2 \pi W_1 a_1 \mu f^{R1}(\alpha) + p^{R2}_v 2 \pi W_2 a_2 \mu f^{R2}(\alpha),
\end{equation}

where $A_P$ is the projected area of Chariklo. The parameter p$_v$ is the geometric albedo of the main body, p$^{Ri}_v$ are the geometric albedo of the rings, $W_i$ are the radial width of the rings, $a_i$ are the radial distance of the rings to the main body, and $\mu$ is the absolute value of the cosine of the aspect angle as seen from Earth, while the phase functions $f(\alpha)$ for the main body and rings are assumed to be equal to unity. This is a good approximation because the phase slope parameter of Chariklo and other centaurs is very small, which is only around 0.06mag/degree, and the phase angle variation of Chariklo is only around 3$^o$.
The parameters $W_i$ and $a_i$ were taken from fits to the occultation
profiles by \cite{Braga2014}. The geometric albedo of Chariklo $p_V$ was
taken from \cite{Fornasier2013}.  We need to mention here that the
\cite{Fornasier2013} observations with the Herschel Space Telescope were
taken during August 2010, where the aspect angle was close to 50$^o$, and
the area of the ring over the area of Chariklo was 0.08. In the albedo determination,
they use $H_{mag}$= 7.4$\pm$0.25, which was transformed into an albedo with a
conservative error that covers the selected $H_{mag}$ for that date.
The obtained albedo with its error was used in our modeling. 

The free parameters of the model are the area of Chariklo and the albedo of the rings. The parameter $A_p$ depends on the ring pole orientation and on the shape of the body. For the spin axis orientation, we used the one preferred given in \cite{Braga2014} that gave satisfactory
results in agreement with the photometry. These values correspond to
$\lambda=138^o$, $\beta=28^o$. The nominal shape used is an ellipsoid, whose projected
shape matches that observed in the stellar occultation.

Using the pole determination and size estimated from the stellar
occultations in Chariklo, we applied the equations presented in
\cite{Tegler2005} to obtain the evolution of the aspect angle, and applied eq. 1
to get absolute magnitudes. The first results that can be seen on these
plots is when the ring plane is crossing the observers (equatorial view or aspect angle = 90$^o$).
This happens in 2007-2008 for Chariklo, as can be seen in figure 1. The other pole solution gives results that are not compatible with the photometry. 

The evolution of the aspect angle can be seen in figure 1 from 1980 to 2030. The vertical lines represents when
the different reflectance spectra were taken (table 2).  The latest spectrum obtained with
X-Shooter in August 2013 has an aspect angle of 58$^o$, which is close to the \cite{BrownMichael1998}
spectra on 1998.  We also plot the area ratio between rings and Chariklo in figure 1. It
is evident that the best time to only see the contribution of Chariklo, when the rings were edge-on, was during 2007-2008, while all the others have
some exposed area from the rings.

\begin{table}
\caption{Visible spectral slope and NIR absorption band depth for all the spectra obtained in literature and the one presented here.}             
\label{table:1}      
\centering                          
\begin{tabular}{l c c c c}        
\hline
\multicolumn{5}{c}{Visible spectra}\\
\hline
date &  slope   &   sigma & aspect angle &reference\\
\hline
1998-03  &   3.82  &    0.45 & 142 &(1)\\
2007-03  &  7.27 &    0.73  &91&(2)\\
2008-02    &  10.36 &    1.04 &83&(3) \\
2013-06&     9.38  &  0.94 &58& this work\\

\hline
\multicolumn{5}{c}{Near-infrared spectra}\\
\hline
date           &  depth [\%]  &    sigma & aspect angle  &reference\\
\hline
2013-06& 11.40 &    8.57  & 58& (this work) \\
2008-02  &  3.97 &   2.50 & 83 &(3)\\
2007-03  &   1.36  &   2.68 & 91&(2)\\
2002-03  &   4.40  &   1.84 & 125& (4)\\
2001-04  &  15.24  &   2.86  & 132& (4) \\
1998-03  &   7.67  &   3.59  & 142&(5)\\
1997-10  &   6.97   &  4.84  & 144 &(6)\\
\hline

\end{tabular}
\tablebib{
     (1)\citet{Barucci1999}; (2)\citet{Alvarez2008}; (3)\citet{Guilbert2009a};
     (4)\citet{Dotto2003}; (5) \citet{BrownMichael1998}; (6)\citet{BrownRobert1998}
}
\end{table}

\subsection{Spectral modeling}

As mentioned before, there are several spectra of Chariklo in the literature
\citep[and references therein]{Guilbert2011}. These spectra show some
similarities and differences that have been interpreted as surface
heterogeneity on Chariklo, as the ring system was not known. All the visible spectra are red, which suggests the
presence of some organics and/or silicates on the surface. Almost all the
NIR spectra show absorptions that are very similar to those of water ice at neither 1.52 and/or 2.02
$\mu$m. However, one of the spectra does not show any trace of water ice
absorption at 1.52 nor at 2.02 $\mu$m. 

Our goal in this section is to study if  the variation on the spectra of Chariklo can be explained based on the
different exposure of the rings, depending on the aspect ratio instead of
based on compositional heterogeneity on the surface of the centaur.

We used the Shkuratov theory \citep{Shkuratov1999} to generate a collection
of synthetic spectra that reproduce the overall shape of the spectrum of the
whole system, which includes Chariklo and rings. These models use the optical constants as input, the relative abundance and the size of the particles of different materials to compute the geometric albedo of the surface at different
wavelengths. These approximations have been widely used to interpret the surface composition of minor icy objects in the Solar System \citep{Pinilla2007, Merlin2010, Poulet2002}. In spite of their impressive history of success, ambiguous results are derived in some cases and have to be taken with care, as none of the solutions are unique.

To reduce this ambiguity, we put special emphasis on the reproduction of the two most
representative characteristics of these spectra: the red slope from the
visible to the NIR and the two water ice bands. To do that, we chose spectra
that cover visible and NIR, or at least J, H, and K bands. If the data at
different wavelengths come from different observations, we chose only those
that are close in time so that the overall shape of the spectrum is not
affected by the variation in the aspect angle. With this approach, we have tried to minimize 
the possibility of misinterpretation.

In the rest of the section, we follow this approach:
First, we model the surface of Chariklo using the spectra that has the smallest contribution from the rings. Second, we constrain the composition of the rings on different dates to show that the origin of the spectral variability is the geometric 
configuration of the system, Chariklo and rings.

\subsubsection*{Modeling of the surface of Chariklo}

As we want to constrain the composition of the surface of Chariklo, we need a spectrum with a minimum contribution from the rings. 
The first spectrum that we modeled was that from \cite{Guilbert2009a}. which was acquired in 2007. According to our photometric model, the aspect angle was 90.4$^o$, as can be seen on figure 1. At this time, almost all the light was
scattered by Chariklo's surface, as the rings were seen by the edge and they are probably very thin. Based on our previous works
(e.g. \citealt{Licandro2005}; \citealt{Lorenzi2014}) we did some tests to
reproduce the shape of the spectrum with materials that are typically found
on the surface of centaurs and trans-Neptunian objects (TNOs): water ice,
complex organics (Triton, Titan, and ice tholins), amorphous olivine, and
amorphous pyroxene with different amounts of magnesium and iron, and
amorphous carbon. 

Finally, as some of the spectra from Chariklo suggests the presence of water ice on the
surface of the main body, we decided to also include water ice in the mixture.
The details of the modeling and the references for the optical constants of
the different materials are shown in table 3.

We also used the estimated value of the albedo of Chariklo in
the visible to constrain the number of solutions. We only selected the models
with an albedo at 0.55 $\mu$m between 3.5\% and 4.1\%  (according to the published estimations
of the albedo \citep{Fornasier2013}). For the evaluation of
the $\chi^{2}$, we normalize the spectrum to the value of the model at 0.55
$\mu$m.

The final number of components in the mixture were determined by the fit. To
chose the best fits we ranked the results by using a $\chi^{2}$ test. From the grid of models, we selected a sample of 30, these are the models 
that  are statistically equivalent with a 90\% confidence \citep{Press1992} according to the value of $\chi^{2}$.
Figure 3 shows the one with the lowest $\chi^{2}$ value that we call the best fit (see table 4).
We can also see the range of variation of these 30 models represented by the gray area around the fit.
For each fit remember that the spectrum is normalized to the value of the
albedo at 0.55 $\mu$m, and we allow some range of variation for the
albedo based on thermo-physical estimations of its value. 

In the next steps, we use this set of models for Chariklo as the canonical
composition of the surface of the centaur without the influence of the
rings; this is what we call thereafter the {\it model of the
continuum}.

\begin{table}
\caption{Characteristics of the models of the continuum.}             
\label{table:1}      
\centering                          
\begin{tabular}{c c c c c}        
material     & abundance (\%)   & size ($\mu$m) &reference\\
\hline
\hline
water ice & $0 - 100$    & 5 - 80 & (1)\\
triton tholin & $0 - 100$   &  5 - 80&(2)\\
olivine  & $0-100$  & 10 - 160      &(3) \\
pyroxene     & $0 - 100$    &10 - 160       & (3)\\
amorphous carbon     & $0 - 100$    & fixed 100&(4) \\
\hline
\end{tabular}
\tablebib{
     (1)\citet{Warren1984}; (2)\citet{McDonald1994}(3)\citet{Dorschner1995};(4)\citet{Rouleau1991}.
}
\end{table}

\begin{figure}
   \centering
   \includegraphics[width=9cm]{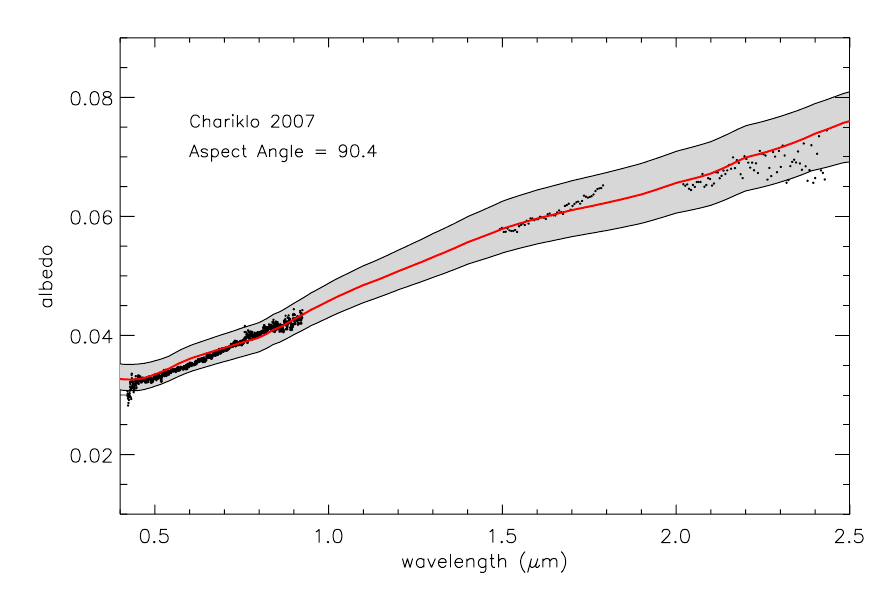}
   \caption{Spectral model of the system Chariklo and rings with the rings edge-on. The red line represents the best-fit to the spectrum (see table 4 for details). The shaded area represents a set of the 30 best fits to the spectrum. The relative reflectance is normalized to the albedo value of each model at 0.55 $\mu$m for the fit; this albedo value is allowed to vary between 0.035 and 0.041 (see text for details).}
   \label{Fig3}
\end{figure}

\subsubsection*{Modeling of the rings}

Now that we have the composition of the surface of Chariklo, the aim is to model the composition of the rings. 
In a second step, we modeled the new spectrum presented in this work. This
particular spectrum has the best characteristics for the modeling, as it was
acquired from 0.5 to 2.3 $\mu$m in only one shot using X-Shooter. It is also a good choice because it was acquired at an aspect angle of 58 $^{o}$ showing clear traces of water ice. For the modeling of this spectrum, we used ``areal mixtures'' of the {\it model of the
continuum} with a grid of intimate models that are created in step one.

In the areal mixture, we used the aspect angle of the rings to calculate the
relative contribution of Chariklo and the rings to the total scattered light
in the spectrum.

The final albedo from the model is\\

A$_{Ch}$ . {\it (model continuum)} + A$_{rings}$ . {\it (model rings)},\\

where A$_{Ch}$ and A$_{rings}$ are the relative normalized area of the centaur and the rings, respectively. The value when A$_{rings}$ = 1 is viewing the ring face-on. This area ratio is taken from the photometric model. For example, for this particular case,  the spectrum of 2013, the contribution from the rings, and centaur are 0.144/0.853, respectively.

There are two different estimations of the albedo of Chariklo in the literature: \cite{Fornasier2013} have found a value of 3.50$\pm$1.1\%, while
\cite{Stansberry2008} have estimated a value of 5.73\%$^{+0.49}_{-0.42}$. We know now that the differences between these values are influenced by the different geometry of the Chariklo and rings system. Different geometry causes a variation in the absolute magnitude, which is an important factor in the determination of the albedo. For this reason, we allow the albedo to vary over all the range from 3.5 to 6.4\% in our models. 
Finally, we ranked the best models using a $\chi^{2}$ test, and we select the collection of models that are statistically equivalent. In figure 4, we show an example of how the modeling works. The details of the best model are shown in table 4. All of our best models for the rings include 20\% of water ice and an 80\% of other materials that are intimately mixed to give the spectrum its overall reddish appearance. We discuss these results in detail in the next section.

\begin{figure*}
    \centering
    \includegraphics[width=18cm]{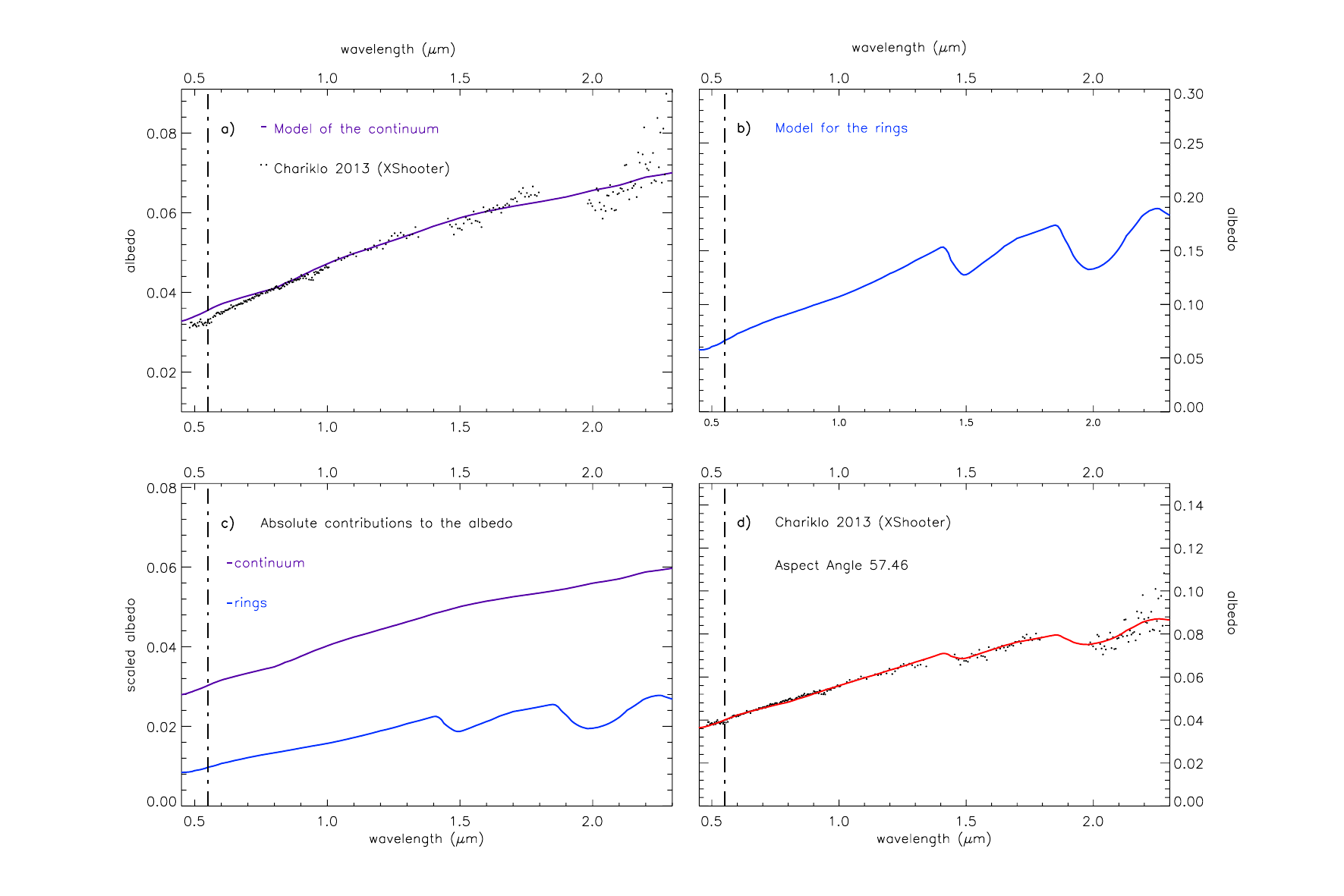}
    \caption{Modeling of the spectrum of Chariklo and rings in 2013:  a) Model of the continuum superimposed to the observed spectra in 2013; b) Model for the rings; c) Absolute contribution of each part, continuum and rings; d) Final model for the spectrum of the system in 2013.}
        \label{Fig4}
\end{figure*}

As a final test, we study if the presence of crystalline water ice, instead of amorphous water ice, 
could improve the results. We found that the value of the $\chi^{2}$ is slightly better when we use crystalline water ice, but the 
improvement is not significant. These tells us that we cannot discard the presence of crystalline water ice due to the signal to noise of the spectrum.

In the last step, we modeled other spectra from the literature. From the
modeling presented above, it is obvious that the overall shape of the
spectrum from 0.5 up to 2.3 $\mu$m is very important. Based on the spectral coverage, not all the spectra
of Chariklo in the literature are good for this effort. We chose spectra with the best spectral coverage in the visible and near infrared and/or simultaneous in time. First, we chose the spectrum from \cite{Dotto2003}, as obtained in
2002. This spectrum was acquired using the high-throughput low resolution
mode of the Near-Infrared Camera and Spectrometer (NICS) at the Telescopio
Nazionale Galileo \citep{Baffa2001}; this instrument incorporates an Amici
prism disperser that yields a $0.8-2.5\ \mu m$ spectrum in only one shot. It provides a reliable shape of the continuum over a broad part of the
wavelength range of our interest. The aspect angle was 125$^o$, showing a good percentage of the total area of the rings.
 We also model the spectrum presented in \cite{Guilbert2009b}  that was obtained in 2008 and presented an aspect angle of 83 $^o$. This spectrum combines two
sets of data acquired in February 2008. One set covers the visible
wavelengths from 0.4 up to 1.0 $\mu$m, while the other covers H and K band
in the NIR. These data were merged together using photometry at V, R, I, J,
H, and K bands (see \cite{Guilbert2009b} for details).
All this procedure results in an spectrum that is adequate for the modeling effort. In 2008 the ratio of the area of the rings and Chariklo exposed was 0.03/0.97, respectively, while it was 0.151/0.849 in 2002. The aspect angle for 2008 was 83$^o$ and 125$^o$ for 2002.

We used again areal mixtures of the {\it model of the continuum} and the same
grid of models that we used for the composition of the rings. We selected the
best model, as we did before. These models are shown in figure 5 with the models for 2007 (continuum) and 2013 for comparison. They are
shifted in the vertical for clarity. The details of the models are included
in table 4.

\begin{figure}
    \centering
    \includegraphics[width=9cm]{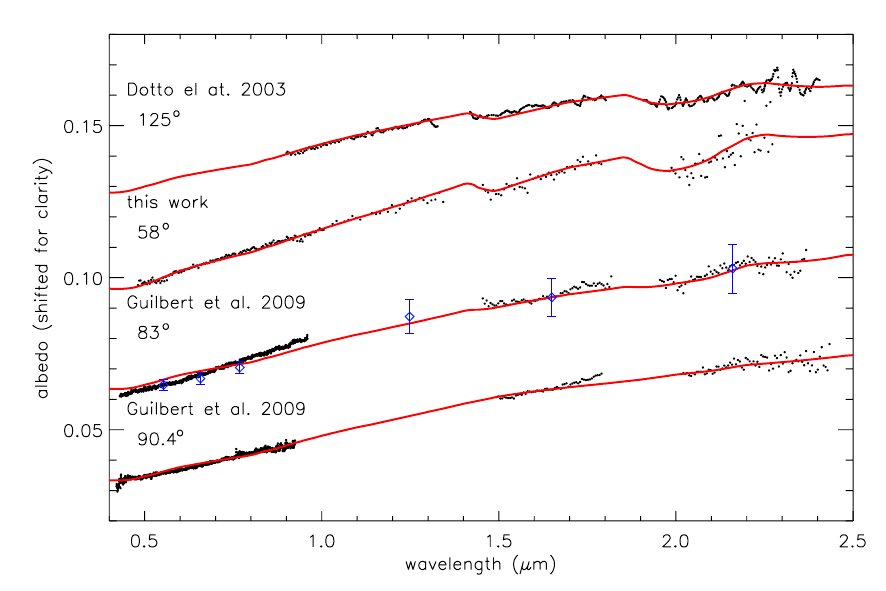}
    \caption{ Models for a set of spectra showing variability. Parameters are shown in table 4. Each spectrum is labeled with the reference and the aspect angle of the system of rings at the date of the observations.
         }
        \label{Fig5}
\end{figure}

\subsubsection*{Results from the modeling}

In table 4, we show the best model selected from our fits to the spectra of Chariklo and its rings at different aspect angles. These were selected using a $\chi^{2}$ criterium. However, these models may not be unique in its goodness of fit, as they rely on the assumptions inherent to the Skhuratov formalism, and on the materials present in the modeling. Each observed spectrum can be fitted by a collection of models that are statistically equivalent. The absence of features in the reflectance of some surface components increases the degree of degeneracy of the solutions. However, these materials are necessary to fit the overall red slope of the continuum. Other materials, such as the water ice, are better constrained due to the presence of two absorption bands in the wavelength range of study. Nonetheless, the models selected provide insight into the surface of Chariklo and the rings. In this section, we expose the main results that we get from the modeling effort.

One of the main results in the modeling comes from the spectrum acquired in 2007, when almost all the light is scattered by Chariklo's surface. Our best models all discard the presence of water ice on the surface of the centaur. They also show that this surface is composed of a mixture of amorphous silicates (30\%), red-complex organics (10\%) and amorphous carbon (60\%). The mixture of these materials is what gives the spectrum its reddish featureless appearance.

The second main result is that we can model all the other spectra with a canonical composition for the continuum which is linearly combined with a canonical composition for the rings, but changing their relative contributions by means of the aspect angle of the system. The rings can contain up to a 20\% of water ice (with a size particle of $\sim$ 100 $\mu$m), and the best selected models fit bands in the three spectra with a very different shape from the very shallow in the spectrum that are found from 2008 (aspect angle = 83$^{\circ}$) to those deeper that are found in the spectrum of 2013 (aspect angle = 58$^{\circ}$). These models clearly discard a composition of the rings dominated by water ice.

However, although the composition of Chariklo and the non-icy components in the rings is pretty similar, it also shows some differences that are important to note. The most evident is the abundance of amorphous carbon that seems to be much higher on the surface of Chariklo than on the rings. This could be an indication that the ring is younger than the surface of the centaur that exhibits a composition more similar to that of highly processed cometary nuclei \citep{Campins2006}. The rings are supposed to be formed by small particles that have collisions with each other; this process can expose the water, showing a younger surface than the main body.

From the results of the modeling we can also have a broad estimation of the albedo of the rings. Revising the albedo at 0.55 $\mu$m from all the best fits, we estimate a value of  7.0$\pm$1.0\% for the albedo of the rings (figure 4). At the same time, the albedo of Chariklo remains pretty low at $\sim$3.6\% $\pm$1.0\% (figure 4). When we scale these albedos considering the aspect angle,  we get the final contribution to the total albedo of the system Chariklo and rings of $\sim$4.0\% $\pm$1.0\%.

While the size of water ice particles is well constrained, the size of the particles for the other components is more degenerated. This is because there are no absorption bands of these materials in the wavelength range for the study, and all of them can be combined in different ways to produce a featureless reddish and low albedo final spectrum similar to that of Chariklo. 

We need to mention here that the size of the particles that we obtained do not correspond to a detailed description of the size distribution of the particles in the rings. The ring could also be populated with larger chunks of material but covered with fine-grained material with the sizes  that we found in our modeling.

\begin{table*}
\caption{Characteristics of the best models of Chariklo and rings used in the figures. These best models were chosen from the collection of models that are statistically equivalent with a 90\% confidence.}             
\label{table:1}      
\centering                          
\begin{tabular}{l c c c c c c| c c c c c c }        
\hline
\hline
 & \multicolumn{5}{c}{Concentration (\%)} &\multicolumn{5}{c}{Particle Size $(\mu$m)}  \\
\hline
year& &Water Ice & Tholin & Olivine & Pyroxene & Carbon & Water Ice & Tholin & Olivine & Pyroxenes & Carbon\\
\hline\hline
2007&&0&10&20&10&60&5&10&210&60&100\\
\hline\hline
2013 & cont &0&10&10&20&60&5&10&110&110&100\\
&ring&20&10&30&40&0&70&25&110&10&100\\
\hline
2008 & cont &0&10&10&20&60&5&10&110&110&100\\
&ring&20&30&10&30&10&90&70&110&10&100\\
\hline
2002 & cont &0&10&10&10&70&5&15&110&210&100\\
& ring &20&20&50&0&10&60&13&210&10&100\\

\hline
\end{tabular}
\end{table*}

\section{Discussion}

The key point in all the results presented in our work is the determination of the spin axis of Chariklo. This was done from the stellar occultation, which is reported by \cite{Braga2014} and assumes that the rings are in the equatorial plane of Chariklo. 
The spin axis orientation which can explain the absolute magnitudes observed
in the 1997-2013 period and the spectroscopy as
already stated in the introduction, is consistent with one of the two
spin axis orientations derived from the stellar occultation reported by
\cite{Braga2014}. However, this orientation also explains the lack of detection
of a rotational lightcurve in the 1997-2000 time frame \cite[e.g.][]{Peixinho2001}, because the aspect angle at that time was nearly 140 degrees. 
Thus a triaxial ellipsoid, a very irregular body, or a body with
albedo variegations on its surface would produce nearly flat, rotational
light curves. \cite{Belskaya2010} already pointed out this possibility and
provided a tentative spin axis orientation that turned out to be off by
around 30 degrees, which is remarkable given the information available at
that time.

The spin axis orientation in our model can also explain why a rotational
lightcurve has been detected in 2013 from high signal-to-noise images with
an amplitude of 0.1mag and a period of 7.004$\pm$0.036 hours \citep{Fornasier2014} .
As shown in fig. 1, the aspect angle in 2013 is large enough to allow a
revealing rotational variability. A triaxial ellipsoid with axes dimensions
a$\sim$122 km, b$\sim$122 km, and c$\sim$117km can reproduce the observed
rotational amplitude in 2013 and the non-detection in 1997-2010, if all the
variability is due to shape.

The correct spin axis orientation not only must explain the rotational lightcurve of Chariklo and the absolute magnitude variability but also the spectral behaviour over time. Our modeling shows that we can explain the changes in the spectrum of Chariklo, and in particular, the shape of the bands of water-ice by means of changes in the aspect angle of a system that is formed by a main body and a system of rings. By modeling the spectrum acquired in 2007, we also discard the presence of water ice on the surface of Chariklo. Moreover, on this date when the rings were edge-on, we did not see any trace of water ice in the spectrum from the system suggesting that the rings must be thin.

This shape of Chariklo is not far from a Maclaurin ellipsoid of density
$\sim$750 kg/m$^3$ which spins at 8h, using the equilibrium figure formalism
(e.g. \cite{Chandrasekhar1987}. This would mean that Chariklo would not be far from
hydrostatic equilibrium, and the derived density seems consistent with that
of small TNOs. From the analysis of rotational
lightcurves of a sample of TNOs and Centaurs, \cite{Duffard2009} concluded
that most TNOs seemed to be in hydrostatic equilibrium and that this
possibility also holds for the Centaurs, at least for the largest ones, such
as Chariklo. A detailed shape from rotational lightcurves and the consideration of the occultation chords has been inverted by Carry et al. (private comunication), although all the variability is supposed to arise from shape in that model,
which is not necessarily the case. The model does not include the effect
of a ring, which slightly decreases the amplitude of the rotational
lightcurve when the aspect angle is not 90$^o$.

Rings are common around the giant planets in the Solar System; however, their appearance and composition are
very different, which probably results from a different origin and the exposition to resurfacing mechanism that acts at different degrees of intensity \cite[][for a review]{Tiscareno2013}. This translates from rings formed by meter-centimeter size particles of almost pure water ice to optically thin rings formed by micrometer-size particles of dust (silicates). The results of our modeling show that the rings of Chariklo are composed of a mixture of bright and dark material. The composition of the rings is more similar to what we would expect from the disruption of an object formed in the same region as Chariklo itself with a mixture of ices, silicates and carbonaceous materials processed by high-energy irradiation. While part of the appearance of the rings of the giant planets is sculpted by the action of the electromagnetic forces, which are generated by the planet itself, in the case of Chariklo, we expect that most of the changes are produced by the action of micro-collisions. The area of Chariklo's rings is about 25 000 km$^{2}$, and as it happens for the planetary rings, we can expect that debris are continuously impacting the material on the ring by causing changes in their orbit and composition. These continuous collisions would sublimate part of the ice exposed at the moment of the formation of the rings. They are also responsible for the fragmentation of the most fragile materials. Another important factor affecting the chemical composition of the grains is the irradiation by high-energy particles that process the ices and cause the amorphization of water ice and the transformation of simple hydrocarbons into complex dark material as the carbon. 

Due to the signal-to-noise of our spectra, we cannot distinguish the state of the water ice neither in the shape of the 1.5 $\mu$m band nor from the modeling. Amorphous water ice is expected due to irradiation, but also crystalline water ice is expected due to the collisions of the fragments in the rings and to the micro-bombardment. We think that the ice is probably a mixture of the crystalline and amorphous state, as it has been inferred in other bodies in the outer solar system \citep{Pinilla2009}.

\section{Conclusions}

Using information on the pole of the rings from the 2013 stellar occultation, we determine the following:

\begin{itemize}

\item We prove that only one of the pole solutions found in \cite{Braga2014} is compatible with the photometry and spectroscopy. 

\item We determined the times when the rings were edge-on and made the connection with the disappearance of the absorption band due to water ice on the spectra of Chariklo.

\item We modeled the composition of Chariklo's surface that is 60\% carbon, 30\% silicates, 10\% organics and no water ice. We also model the composition of the rings that is 20\% water ice, 10 - 30\% organics, and 40-70\% silicates and small quantities of carbon. With this modeling, we discard the presence of water ice on the surface of Chariklo independently of the date of the observation. We confirm that the spectroscopic variation is due to a different area exposition of the rings which is not a variation of the surface of Chariklo itself.

\item  Except for the water, the composition of the rings is compatible with the same materials as the surface of Chariklo but combined in different relative abundances: Amorphous Carbon from 0-20\%, amorphous silicates from 40-70\%, and organics from 10-30\%.

\item The amount of water ice is well constrained by the models due to the presence of two absorption bands in the wavelength of the study and has to be around 20\% with a particle size $<$100 $\mu$m. As soon as the aspect angle is different from 90$^{o}$, the water ice band is visible again by showing that a system of rings, such as the one detected for Chariklo wiht 20\% of water ice and as the one infered from our models, is evident in the NIR spectra at deviations of 10$^{o}$ or higher from the edge-on configuration.

\end{itemize}

\begin{acknowledgements}
RD acknowledge the support of MINECO for his Ram\'on y Cajal Contract. AAC thanks CNPq and FAPERJ for financial support. BS acknowledges support from the French grant ANR-11-IS56-0002 "Beyond Neptune II". FBR  acknowledge the support of CNPq, Brazil  (grant 150541/2013-9). Funding from Spanish grant AYA-2011-30106-CO2-O1 is acknowledged, as well as the Proyecto de Excelencia de la Junta de Andalucía, J.A.2007-FQM2998 and FEDER funds. 
\end{acknowledgements}


\bibliographystyle{aa}
\bibliography{centaurs}
\end{document}